\documentclass[final,5p]{elsarticle}
\usepackage{graphicx,amssymb,amsmath,hyperref,xspace,tabularx,pifont,floatrow}

\def\si{$^{32}$Si\xspace}
\def\p{$^{32}$P\xspace}
\def\esi{$^{28}$Si\xspace}

\journal{Astroparticle Physics}

\biboptions{numbers,sort&compress}
\begin{document}

\begin{frontmatter}

\title{\boldmath Naturally occurring \si and low-background silicon dark matter detectors}

\author[]{John L. Orrell}
\author[]{Isaac J. Arnquist}
\author[]{Mary Bliss}
\author[]{Raymond Bunker\corref{rab}}\ead{Raymond.Bunker@pnnl.gov}
\author[]{Zachary S. Finch}
\address{Pacific Northwest National Laboratory, Richland, WA 99352, U.S.A.}
\cortext[rab]{Corresponding author}

\begin{abstract}
The naturally occurring radioisotope \si represents a potentially limiting background in future dark matter direct-detection experiments. We investigate sources of \si and the vectors by which it comes to reside in silicon crystals used for fabrication of radiation detectors. We infer that the \si concentration in commercial single-crystal silicon is likely variable, dependent upon the specific geologic and hydrologic history of the source (or sources) of silicon ``ore'' and the details of the silicon-refinement process. The silicon production industry is large, highly segmented by refining step, and multifaceted in terms of final product type, from which we conclude that production of \si-mitigated crystals requires \textit{both} targeted silicon material selection \textit{and} a dedicated refinement-through-crystal-production process. We review options for source material selection, including quartz from an underground source and silicon isotopically reduced in \si. To quantitatively evaluate the \si content in silicon metal and precursor materials, we propose analytic methods employing chemical processing and radiometric measurements. Ultimately, it appears feasible to produce silicon detectors with low levels of \si, though significant assay method development is required to validate this claim and thereby enable a quality assurance program during an actual controlled silicon-detector production cycle.
\end{abstract}


\end{frontmatter}

\section{Introduction}\label{sec:intro}
High-purity silicon crystals are used as detector substrates in dark matter direct-detection experiments such as CDMS II \cite{cdms2_si} and DAMIC \cite{damic}. Silicon crystals are instrumented with ionization and/or phonon sensors to measure energy deposited by theorized galactic dark matter particles scattering off silicon nuclei \cite{lewin_smith}. The small cross section inherent in the dark-matter--nucleon interaction --- of order 10$^{-41}$ cm$^2$ or smaller for dark matter particle  masses of 5 GeV/$c^2$ --- suggests very low event rates for kilogram-sized detectors (events per year), requiring a concomitant low rate of interactions from potentially confounding sources of background radiation. Consequently, radioisotopes transported through the environment and into detectors, or cosmogenically produced directly in detectors, are a significant concern for future generations of low-background experiments (see, e.g., Refs.\ \cite{ARMENGAUD201751,cresst,nai}).

Following the observation of \si in detector-grade silicon \cite{damic_backgrounds}, the background from \si decays has emerged as a particular concern and has the potential to limit the dark matter sensitivity of future silicon-based detectors.
Although the betas emitted in decays of \si (and its \p daughter) interact with a detector's orbital electrons, as opposed to the nuclear recoils expected from dark matter interactions, the resulting signals measured by a detector may nevertheless mimic or overshadow the sought after dark matter events if the \si decay rate is too large. Energy depositions from dark matter interactions, in particular for low-mass dark matter particles with few-GeV/$c^2$ masses (or less), are expected to be at the keV-scale (or less) \cite{ge_diode}. Direct-detection techniques that rely on differences between electronic and nuclear recoils to discriminate background interactions tend to break down at such low energies (cf.\ Fig.\ 9 in Ref.\ \cite{cdms_shallow}), because there is typically little to no additional event information beyond the statistically significant observation that the events occurred just above the detection threshold.

In this article, we explore sources of \si and evaluate its injection vectors into the production cycle for high-purity, single-crystal silicon.
Two general vectors are considered: the initial silicon raw material, and introduction during purification and crystal-production processes. Within this context, the commercial silicon production chain is evaluated and a potential alternative source of isotopically pure silicon is highlighted. In both cases, there is significant planning and fabrication value in developing methods for determining \si levels prior to crystal growth and detector fabrication. Such assay methods would help ensure that future low-background detectors contain sufficiently low levels of \si. To this end we propose assay method concepts employing chemical-separation techniques to isolate the daughter \p atoms for use in radiometric counting analysis. 

Finally, we outline potentially viable approaches to obtaining, validating, and producing \si-mitigated crystals. A benchmark for future dark matter experiments is detection of solar-neutrino coherent scattering on silicon nuclei, corresponding to a dark-matter--nucleon cross section $<$\,10$^{-44}$\,cm$^{2}$ at few-GeV/$c^2$ masses. To evaluate the feasibility of our proposals, we present a simple calculation comparing the rate of solar-neutrino coherent scattering to the rate of \si and \p beta decays normalized to different \si concentrations. The nominal goal is for the latter to correspond to a decay rate (per kg Si) $\sim$10$\times$ lower than the former and thus allow for exploration of dark matter interaction rates as low as those expected from solar-neutrino-induced nuclear recoils. 

\section{Silicon-32 in detectors}\label{sec:detectors}
Early evaluations of the use of silicon detectors for measurement of rare-event processes  suggested that natural radioactivity inherent in high-chemical-purity silicon crystals would not hinder such physics experiments.  This notion is supported by the lack of any significant (direct) production channels for cosmogenic activation of long-lived silicon radioisotopes (i.e.\ \si) in natural silicon, unlike many other detector materials of interest to the astroparticle physics community (e.g.\ $^{68}$Ge in germanium crystals and $^{39}$Ar in argon gas~\cite{cebrian}). In Ref.\,\cite{martoff}, e.g.,  Martoff notes that silicate rock sourced from deep underground should be low in \si because it has been isolated from the atmospheric \si source for a sufficiently long time. A later review by Plaga acknowledges \si as a potential background issue and advises  use of silicate rock that has not mixed with ground water as a source material for low-background silicon detectors \cite{plaga}. Plaga goes on to cite measurements in which water from deep wells showed \si activities of 0.04--0.06 decays/(kg\,d) \cite{ground_water}, thereby cautioning that deep underground water may not be entirely free of \si; thus it is not certain that underground sources of silicate rock will necessarily be free of \si. A subsequent review mentions the concern of \si in silicon-based detectors, but does not elaborate \cite{heusser}. These early reviews hint at the connection to the underlying contamination mechanism --- atmospheric spallation of $^{40}$Ar and transport into source material via precipitation --- but they provide few details to address the problem in practical detector-fabrication planning.

Two experiments that used silicon detectors to search for dark matter have reported that there was in fact \si in their high-purity detector substrates. Specifically, Caldwell et al.\ suggest that the low-background spectrum measured with their array of silicon detectors is consistent with a \si activity of $\sim$300 decays/(kg\,d) \cite{cosmion}, from which they conclude that their crystals were likely fabricated from surface sand that had mixed freely with \si precipitated from the atmosphere. More recently, the DAMIC collaboration reported a rate of 80$^{+110}_{-65}$ decays of \si per day (at 95\% C.L.)\ in each kg of their silicon detector material \cite{damic_backgrounds}. Their analysis is particularly convincing because they were able to localize the \si beta in their CCD detectors and correlate the subsequent energy deposition from the daughter \p beta, both spatially and temporally (cf.~Fig.\,8 in Ref.\ \cite{damic_backgrounds}).  

These measurements highlight the need for a better understanding of the \si levels expected in future dark matter detectors. In particular, the SuperCDMS SNOLAB experiment will perform a low-mass ($<$\,10 GeV/$c^2$) dark matter search employing approximately 3.6 kg of silicon detectors \cite{sensitivity}. A sensitivity study shows that the dark-matter--nucleon cross-section reach of these detectors will be limited by \si decays if present at the level reported in the DAMIC CCDs (cf.\ Fig.\ 9 in Ref.\ \cite{sensitivity}). Measurement and reduction of \si at (and below) the DAMIC level is a likely prerequisite for any future experiment that attempts to push beyond the anticipated sensitivity of SuperCDMS SNOLAB using silicon detectors.

\section{Silicon-32 in the environment}\label{sec:environment}
The presence of \si in silicon-detector source material (or ``ore'') is due to its natural occurrence in the environment \cite{lal}. In Earth's atmosphere, only the noble gases argon, krypton, and xenon have nuclei with $A$ and $Z$ values greater than \si. Argon isotopes --- principally $^{40}$Ar at 99.6\% natural abundance and at a concentration $\sim$7500$\times$ higher than the other two noble gases combined --- are the primary target nuclei for cosmic-ray spallation processes that generate \si\ \cite{lal2}.\footnote{In principle, cosmic-ray secondary particles may produce \si through spallation processes via direct interaction with nuclei in solid terrestrial materials  (e.g.\ rock and dirt). However, \si is not typically suggested as a radio-tracer from terrestrial production and does not appear to have been studied along with the five radionuclides commonly used for tracing materials of terrestrial origin: $^{10}$Be, $^{14}$C, $^{26}$Al, $^{36}$Cl and $^{39}$Ar \cite{lal5, gosse}. For this reason, we do not consider this potential terrestrial source vector for \si.} Estimates of the total \si production rate suggest an atmospheric global-average surface injection rate of $\sim$10$^{-2}$ atoms/cm$^2$/min \cite{lal3}.  However, a more recent evaluation has questioned the assumptions used for the surface injection model, suggesting a rate $\sim$2$\times$ smaller \cite{paradox}. Additionally, an integrated in situ oceanic production rate of 2.5$\times$10$^{-5}$ atoms/cm$^{2}$/min is principally due to spallation on sulfur and calcium \cite{lal4}.

\begin{figure*}[ht]
\centering
\includegraphics[width=.99\textwidth]{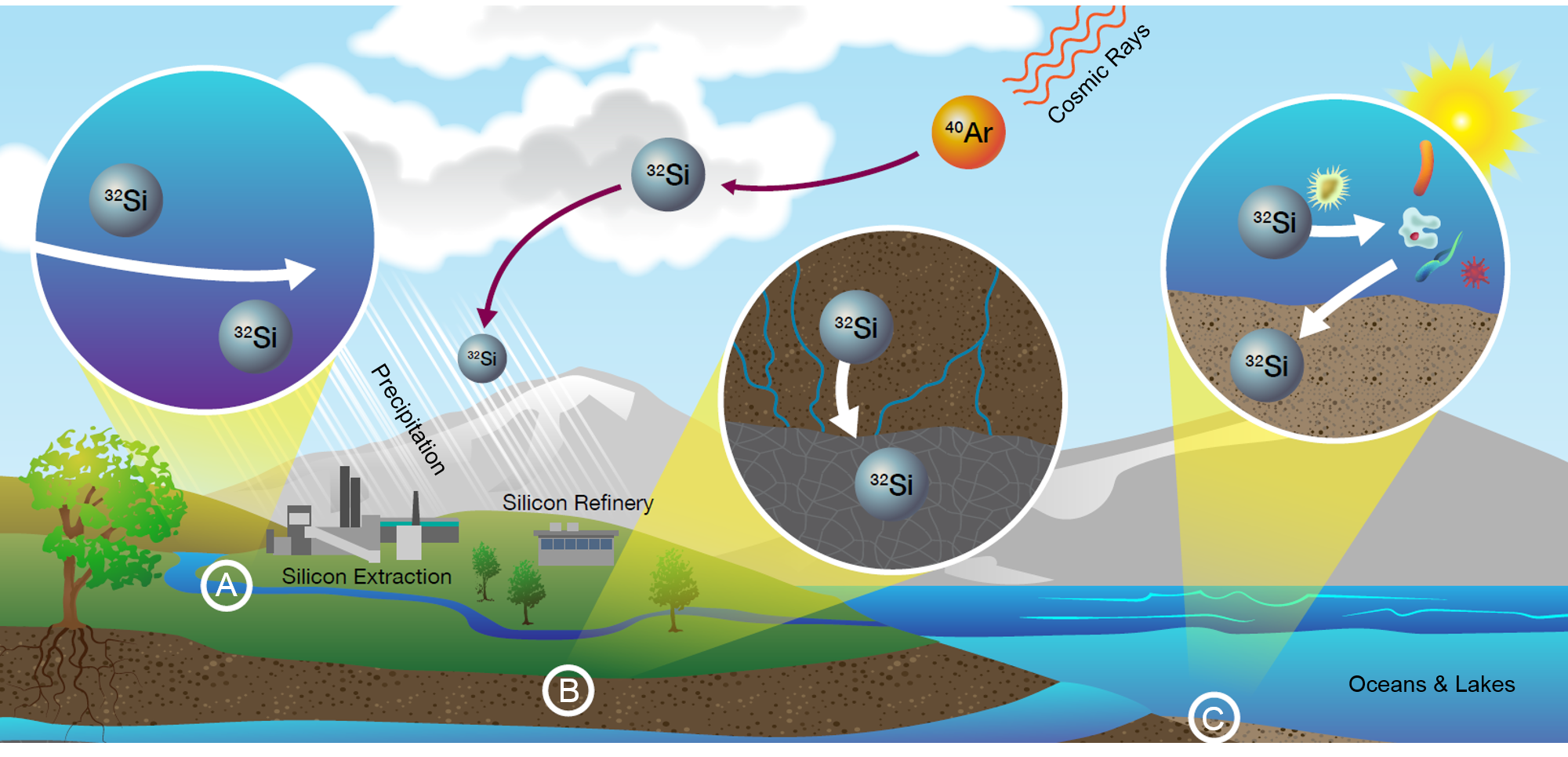}
\caption{\small Conjectured transport and accumulation of cosmogenically created \si in the terrestrial environment. Cosmic rays interact with $^{40}$Ar in the atmosphere to spallate \si that is then transported into the terrestrial environment via precipitation, leading to accumulation of \si in:  A) streams and settling ponds that may be sources of processing water for silicon mining and refinement; B) surface sands and near-surface silicon deposits; and C) oceans and lakes, where it can be transported by biological organisms and ultimately incorporated into the underlying sediments.} 
\label{fig:silicon-32}
\end{figure*}

As the vast majority of commodity silicon is produced from mining in the terrestrial environment, the primary considerations for understanding the level of \si present in any future silicon-based radiation detectors are the geologic age of the silicon ore and the previously conjectured \cite{plaga} interaction of the ore with precipitated rain water or ocean water that contains \si scavenged from the atmosphere. Any ore separated from injections of atmospherically produced \si should be of sufficient age (greater than several thousand years) to have little to no \si because of its relatively short (geologically speaking) 153-year half-life \cite{NNDC}. This then focuses attention on the \si injection and accumulation mechanisms for silicon ores. 

As noted in the literature (e.g.\ Chapter 3 in Ref.\ \cite{lal6}), \si is a  non-conservative tracer. Conservative tracers are chemicals that pass through a system without being removed through chemical reactions or adsorptive processes. Studies of silica in aqueous soil solutions show that the dissolved-silica concentration in water evolves over time, implying an exchange process whereby silica is brought into solution and then re-adheres to soil particulates \cite{mckeague}. This observation has clear bearing on the understanding of how cosmogenically produced \si is transported through the environment and comes to reside in silicon ores. A study using $^{31}$Si-labeled silica showed an exchange reaction takes place at the quartz surface via dissolved H$_4$SiO$_4$ (monosilicic acid) \cite{holt}, thus supporting the conjecture that \si in precipitated rain water can accumulate in quartz and sand. Figure \ref{fig:silicon-32} is suggestive of the transport mechanisms underlying this accumulation of \si in terrestrially-located materials.

The presence of \si in ore material due to contact with precipitated surface water is a compelling conjecture;  however, it could also be introduced during crystal production as a result of the refinement process. These potential \si injection vectors are discussed in the next section and should be considered together with the above review when seeking means to mitigate or control \si levels in future radiation detectors, either through judicious selection of the raw ore material and/or by controlling the silicon-refinement process.

\section{Commercial silicon production}\label{sec:commercial}
\begin{figure*}[ht]
\centering
\includegraphics[width=0.95\textwidth]{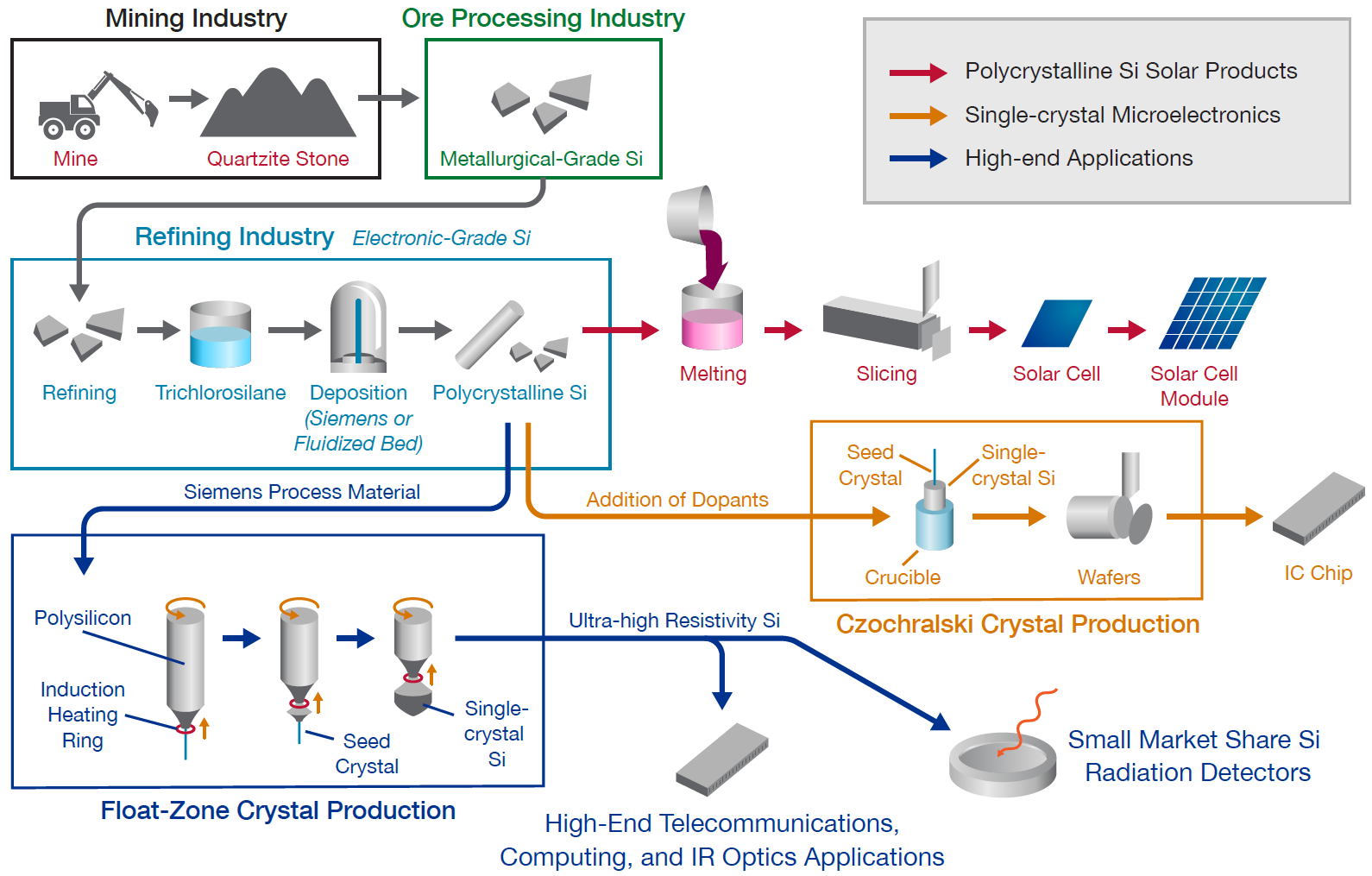}
\caption{\small Simplified outline of the commercial silicon production cycle: from mining of silicon ``ore,'' to metallurgical-grade production, to polysilicon refinement, and finally to single-crystal growth.} 
\label{fig:production}
\end{figure*}

In this section we review the commercial silicon production cycle and supply chains, organized by the production stages and potential \si injection vectors. Generally, we note that the silicon industry is large and highly segmented by processing step: from mining, to metallurgical-grade production, to polycrystalline-silicon refinement, and finally to single-crystal growth (see Fig.\ \ref{fig:production}). Consequently, we conclude that it is likely difficult to establish and maintain provenance of silicon materials throughout the commercial production cycle.

\subsection{Overview of production cycle}\label{ssec:production-cycle}
The various grades of commodity silicon are based on purity. Ferrosilicon alloy represents $\sim$65\% of worldwide production while ``pure'' silicon metal is the remaining $\sim$35\% \cite{usgs_minerals_2017}.\footnote{Percentages are based on silicon content by weight.} The latter is divided into ``metal'' and ``chemical'' grades. Chemical grades are primarily used to produce silicone polymers and silica reagents, whereas metal grades are typically used in aluminum alloys and the semiconductor industry \cite{usgs_website}. The semiconductor industry represents a small fraction of global production and is also divided into purity grades, with purity of the monocrystalline grades measured in terms of resistivity --- a more useful metric for this form because of its relevance to end-product applications. The definition of the various grades varies somewhat, but they are generally categorized by chemical purity (or progressively higher resistivity), as presented in Table \ref{tab:grades}.

\begin{table*}[ht]
\small
\begin{center}
\begin{tabular}{ l r r }
\hline 
 & & \\[-1em]
Category & Chemical Purity & Notes\\
 & & \\[-1em]
\hline
 & & \\[-1em]
Ferrosilicon alloy & 15--90\% & iron-silicon alloy \cite{ferrohandbook} \\
Metallurgical grade & $\sim$98\% & reacted with carbon \\
High-purity metallurgical grade & 3N--6N & ``upgraded'' \cite{metallurgic}\\
Solar grade & 6N--7N & photovoltaics \cite{braga}\\
Fluidized bed reactor polysilicon & 6N--9N & see, e.g., Ref.\ \cite{FBR} \\
Semiconductor grade & $>$9N & microelectronics\\
Siemens process  polysilicon & 9N--11N & chemical vapor deposition\\
& & (cf.\ Fig.\ 6 in Ref.\ \cite{metallurgic}) \\[-0.5em]
\textit{Single crystal} &  &  \\
\quad Czochralski growth & 100 Ohm-cm & silica crucibles, oxygen doping \cite{cz} \\
\quad Magnetic Czochralski growth & 1,000 Ohm-cm & larger crystals, lower oxygen \cite{mcz} \\
\quad Floating-zone growth & $>$10,000 Ohm-cm & crucible-free, highest purity \cite{fz} \\
\hline
\end{tabular}
\end{center}
\caption{\small Grades of commercially produced metallic silicon, categorized by purity and listed from lowest (top) to highest purity (bottom), separately for polysilicon and its precursors (upper rows) and single-crystal silicon (lowest three rows). The ``N'' notation indicates the number of consecutive nines in the purity percentage (e.g.\ 3N $\equiv$ 99.9\%). Purity for the single-crystal categories is measured in terms of resistivity. For reference, the dopant concentration in 10,000\,Ohm-cm single-crystal silicon is generally equivalent to $\sim$11N chemical purity~\cite{hull}.}
\label{tab:grades}
\end{table*}

Over 40 billion metric tons of crushed stone, sand and gravel are mined annually \cite{krausmann}, with $\sim$1 billion metric tons of sand and gravel produced in the U.S.\ \cite{usgs_minerals_2017}. Most of the material is used in the construction industry. River and beach sands are directly ``mined'' for use as silica ore and consist of rounded particles that are the result of prolonged exposure to the eroding effects of water. The primary gravel sources of silicon are quartzite, flint, and chalcedony, which are chemically identical (SiO$_2$) but have differing microstructures. Quartzite particles are sharp and irregular as a result of the metamorphic process by which they form. Quartzite is essentially tectonically compacted beach sand, while the other mineral forms are much finer grained. The semiconductor industry prefers chemically cleaner ores than are generally used in the production of metallurgical-grade silicon, choosing quartzite because it is already $\sim$80\% chemically pure quartz --- crystalline SiO$_2$ in a silicon-oxygen tetrahedron (SiO$_4$) \cite{quartzpage}. Quartzite varies in hardness but generally requires blasting and crushing as part of the mining process.

It is noteworthy that the highest-purity quartzite deposits, e.g.\ from the Spruce Pine Mine in North Carolina, are extracted for refinement into quartz crucibles that are used for the growth of semiconductor crystals. The very high chemical purity of such deposits ensures that the crucibles do not contribute unwanted impurities during the silicon-crystal growth process. 

Silicon ore is transformed into progressively higher-purity silicon metal through a series of refinement processes.  Feedstock for metallic-silicon smelting has strict purity requirements compared to alloy-grade silicon, but both are typically reduced to metal by mixing with carbon in electric arc furnaces.\footnote{See Refs.\,\cite{ferrohandbook,aasly} for detailed descriptions of the smelting process.}  Once the metallic phase is produced, the remaining trace elements may be removed. The most common methods for further refinement utilize chemical vapor deposition. Silicon metal is reacted with HCl gas to form HSiCl$_3$, which has an advantageously low boiling point of 32\,$^{\circ}$C and can therefore be purified to a very high degree via fractional distillation. This gas is then converted back to metallic silicon via reaction with H$_2$ gas. The Siemens process and fluidized bed reactors are the primary re-solidifcation methods for production of solar-grade polysilicon and feedstock for single-crystal growth. The former uses a resistively heated rod or U-shape of silicon. The HSiCl$_3$ gas decomposes on the hot surface, leaving behind the metal and thereby increasing the diameter of the rod. Fluidized bed reactors use pebbles of silicon suspended in a hot gas stream to grow polysilicon. It is a continuous process in which larger pebbles fall out of the gas flow and are collected for further processing. Fluidized bed reactors consume less energy and are viewed as cheaper to operate than Siemens process reactors.   

The methods used for single-crystal growth impact the chemical purity of the resulting silicon. Most semiconductor wafers are grown via the Czochralski method \cite{cz,mcz} in which a crystal is pulled from molten silicon metal contained in a fused-silica crucible. The crucible gradually dissolves and thus contaminates the final crystal with a trace amount of oxygen that serves to passivate electrical defects created by other trace impurities. The oxygen-doped silicon also has better thermal properties for producing multilayer devices. Wafers made from these crystals are primarily used as substrates to support deposition and patterning of thin-film layers.

The floating-zone (FZ) growth method \cite{fz} produces the highest-resistivity single-crystal silicon. A high-purity polycrystalline rod, typically formed from the Siemens process, and a monocrystalline seed are heated via induction and brought together to form a narrow container-less molten region that is supported by surface tension (cf.\ lower-left box in Fig.~\ref{fig:production} labeled ``Float-Zone Crystal Production''). As the molten region is moved along the length of the rod, it carries with it trace (non-silicon) elements and leaves behind a single crystal with higher chemical purity.   It is this higher-purity material that is typically used for creation of radiation detectors to search for dark matter.

We now turn our attention to specific vectors whereby \si may enter the commercial production cycle and supply chains.

\subsection{Vector 1: Si-32 in the source material}\label{ssec:vector1}
From purely geologic considerations, the quartzite ore used as raw material for the growth of semiconductor-grade silicon should have low levels of \si, because it has (generally) been isolated from contact with surface waters for a very long time prior to mining activities. However, in nearly all cases, the pragmatic costs associated with mining and extraction of quartzite deposits lead to surface mines and thus long-term exposure to surface waters. Infiltration of atmospherically precipitated \si into the strata of the quartzite deposits is therefore expected. The extent of the infiltration and the resulting \si accumulation levels will depend on local geologic details specific to the rock formation. As an example, studies of \si concentrations in limestone aquifers show \si transport from the surface down to depths of several tens of meters \cite{MORGENSTERN}. There is no current economic driver for significant quantities of ``deep'' industrial silica or quartz mining that could produce silicon segregated from surface waters carrying \si.

\subsection{Vector 2: Si-32 introduced during mining}\label{ssec:vector2}
Dust mitigation is paramount in regulated mining operations. Silicosis, a condition comparable to black lung disease, is associated with inhalation of silica dust. To mitigate health concerns, OSHA issued revised Permissible Exposure Limits in 2016, along with dust-control requirements for workplace silica controls.\footnote{OSHA is the Occupational Safety and Health Administration within the U.S.\ Department of Labor.  Silica-dust exposure limits are mandated in \textit{Final Rule} 81:16285--16890, ``Occupational Exposure to Respirable Crystalline Silica.''} This regulation prefers engineering controls over the use of respirators to reduce exposure; so water sprays to control dust are common in the mining industry \cite{niosh}. Further, environmental regulations drive processes to retain and recycle water used for dust control; settling ponds and aeration are also common \cite{EPA}. Silica readily forms aqueous compounds, especially in basic solutions. Consequently, recycled water in mining and ore operations is a strong potential vector for introduction of \si into the silicon supply chain. 

Additionally, wet sieving is common in silicon mining to select grain size. The ideal particle size for reduction to metal is 10--150 $\mu$m. Finer particles cause incomplete reduction during smelting and are specifically excluded. The key conclusion from this discussion is that surface water is used throughout the mining process and is in routine contact with the silicon ore. The trace amount of atmospherically precipitated \si in the surface water therefore has ample opportunity to become incorporated into the supply chain at this step in the production cycle.

\subsection{Vector 3: Si-32 introduced during refinement}\label{ssec:vector3}
As noted above in Sec.\ \ref{ssec:production-cycle}, the Siemens process is a widely used method for refinement of silicon metal --- $\sim$90\% of all high-purity silicon according to Ref.\ \cite{metallurgic}. The method relies on conversion to HSiCl$_3$ gas to achieve an exceptionally high degree of purification. In general, chemical refinement involves use and recycling of chlorine compounds such as  trichlorosilane (HSiCl$_3$), hydrochloric acid (HCl), and silicon tetrachloride (SiCl$_4$). Similar to argon isotopes in the atmosphere, chlorine is a viable target for production of \si through cosmic-ray spallation processes. This second cosmogenic-production channel represents an additional \si injection vector; most high-purity polysilicon is susceptible to inclusion of \si due to chemical contact with chlorine during the refinement process.

It is instructive to compare the scale of this chlorine vector to the \si concentration measured by the DAMIC CCDs. This can be done using \textsc{Geant4} \cite{geant,geant_other} to simulate neutron interactions with chlorine.  Cosmic-ray neutron secondaries dominate the production of isotopes via spallation processes when materials are at sea level. Consequently, the \si production rate can be estimated by integrating the product of the differential production cross section --- extracted from a \textsc{Geant4} Monte Carlo simulation of neutrons incident on chlorine\footnote{There is variability in the production cross section when estimated in this way, depending on which \textsc{Geant4} ``physics list'' is used. We chose the QGSP\_INCLXX\_HP list as the option most likely to have the best accuracy for our application \cite{dennis}; it includes the Li\`{e}ge Intra Nuclear Cascade model (INCL$+$$+$) \cite{INCLXX}, for generation of final states resulting from inelastic scattering of neutrons on nuclei, and the high-precision neutron model (HP) \cite{geant_physics}. The resulting \si production cross section in chlorine is $\sim$1 mb for 1 GeV neutrons.} --- and the differential flux of neutrons at sea level \cite{mei,gordon} (cf.\ Fig.~1 in Ref.~\cite{mei}), for neutron energies between $\sim$10 MeV and 10\,GeV.  The production rate reaches a saturation (or equilibrium) level corresponding to a \si activity of a few decays per kg of chlorine per day, or $\sim$1 \si decay/(kg\,d) when stoichiometrically converted to an equivalent mass of silicon.  This corresponds to $\sim$1--10\% of the concentration measured by the DAMIC CCDs, considering their full 95\% confidence interval, and it would take several hundred years of sea-level exposure for even this low level to build up in the chlorine-containing chemicals used during polysilicon refinement.  

Therefore it appears unlikely that this chlorine vector is the dominant mechanism by which \si enters  the production cycle. However, it is worth keeping in mind because it has the potential to become an important contributer if \si can be mitigated (elsewhere) by an order of magnitude or more.  It also suggests specific \si assay targets, both before and after polysilicon refinement, in order to control any variability in the \si injection rate associated with recycling of chlorine compounds.

\subsection{Vector 4: Si-32 introduced during crystal growth}\label{ssec:vector4}
Although not directly related to the production of the type of high-resistivity monocrystalline silicon used for radiation detectors (i.e.\ FZ-grown), for completeness we cover Czochralski crystal growth because it highlights how production materials may introduce \si into single-crystal material.  Further, such material could lead to an indirect contamination of FZ-grown crystals as a result of material recycling.

Czochralski-growth crucibles are made from or lined with fused silica. As noted in Sec.\ \ref{ssec:production-cycle}, crucibles used in the growth of semiconductor crystals are made from the highest-purity ores, which are utilized because chemical purification of SiO$_2$ is difficult and costly. Before it is melted into crucibles, the high-purity quartz is typically purified further using various water-based methods such as froth flotation to remove lower-density contaminants and aqueous acid washing. 
Thus, although the crucibles are chemically pure, they may yet contain trace levels of \si due to the ore, mining, and processing steps described here and in Secs.\ \ref{ssec:vector1}--\ref{ssec:vector3}. Consequently, there is potential for transport of \si into single-crystal silicon from the crucibles or other silicon-containing equipment used in the crystal-growth process. 

Although this \si injection vector seems improbable from simple considerations of thermodynamic mass transport (i.e.\ diffusion) of silicon atoms out of crucibles (or other silicon-containing materials) into the single-crystal structure, any process specifically intended to produce silicon crystals with low levels of \si will need to consider it in more detail than is treated here. 

\subsection{Vector 5: Si-32 introduced during device fabrication}\label{ssec:vector5}
Finally, \si may be introduced through the processes used to fabricate a device following growth of the single-crystal silicon substrate. This injection vector can only result in an increased \si concentration on device surfaces, rather than in the bulk of the solid substrate. Although we focus here on methods used for the fabrication of dark matter detectors, this vector applies generally to any type of device fabrication.

Water and/or reagents used in crystal-substrate shaping and wafering may introduce \si if the water and/or reagents were sourced from surface waters and were not specifically purified to remove trace levels of silicon.  However, the acid, base, and water washing steps used throughout the semiconductor manufacturing process use high-purity solutions and thus are probably not a significant source of \si. Consequently, we consider this fifth vector as unlikely to be problematic during these early stages of device fabrication.

Of potentially greater concern are the thin films deposited onto device surfaces to create readout sensors. Fabrication of CCDs can include chemical vapor deposition of SiO$_{2}$/Si$_{3}$N$_{4}$ dielectric films and polysilicon epitaxial layers (see, e.g., Ref.~\cite{ccd_fab}), all of which may be vectors for \si. Similarly, fabrication of SuperCDMS-style detectors typically includes deposition (via plasma sputtering) of a poly- or amorphous-silicon ``blocking'' layer  that underlies the charge and phonon sensors~\cite{scdms_fab}.

Such a surface-only source would likely have distinctive energy-deposition features in the DAMIC CCDs, in contrast to the analysis presented in Ref.\ \cite{damic_backgrounds} which assumes a bulk contamination of \si. Also, it seems likely that the primary \si-contamination mechanisms are the vectors described in Secs.~\ref{ssec:vector1} and \ref{ssec:vector2}, in which case the source of \si in any silicon-containing thin films is the same as for the single-crystal substrate onto which they are deposited. Consequently, one might expect a similar \si concentration in the thin films as in the bulk substrate; \si in the bulk would therefore be of principle concern because of the substrate's much larger mass. However, this depends on the level of \si variability in silicon products, which may be considerable, and whether \si has been mitigated in the single-crystal substrate (as outlined in Sec.~\ref{sec:implications}). For the latter case in particular, the injection vector described here may need to be considered. 

As an example, consider the SuperCDMS SNOLAB detector design in which the bulk substrate is cylindrical with 100\,mm diameter flat faces and a total silicon mass of $\sim$600\,g~\cite{sensitivity}.  The aforementioned blocking layers --- one on each of the flat faces --- are 40\,nm thick and therefore have a combined Si mass of no more than 1.5\,mg.  If the \si level measured by the DAMIC CCDs is representative of the concentration in the blocking-layer films, their mass-averaged contribution to a detector's overall \si concentration would be $<$0.0002~\si decays per kg Si per day.  This represents a small enough rate that controlling \si in the thin films is likely unnecessary.


\section{Alternative isotopically pure silicon sources}\label{sec:isotopically_pure}
\begin{table*}[ht]
\small
\begin{center}
\begin{tabular}{ l l l l c c c c}
\hline 
& & & & & & \\[-1em]
Material &  \multicolumn{3}{l}{\underline{Isotopic Composition \% (uncertainty)}}   &       & \multicolumn{2}{c}{\underline{Ratio of IUPAC}}\\
& & & & & & \\[-1em]
Identifier & $^{28}$Si & $^{29}$Si & $^{30}$Si & Data Source & to $^{29}$Si & to $^{30}$Si  \\
& & & & & & \\[-1em]
\hline
& & & & & & \\[-1em]
IUPAC & 92.223(19) & 4.685(08) & 3.092(11) & Table 5 in Ref.\ \cite{IUPAC} & 1 & 1 \\
$^{28}$Si (ORNL) & 99.0176(10) & 0.6025(04) & 0.3799(10) & Table II in Ref.\ \cite{becker} & 7.8 & 8.1 \\
$^{28}$Si-10Pr11 & 99.995752(12) & 0.004136(11) & 0.0001121(14) & Table 5 in Ref.\ \cite{fujii} & 1133 & 27583 \\
 $^{28}$Si-23Pr11 & 99.9984416(46) & 0.0014973(45) & 0.00006104(62) & Table 5 in Ref.\ \cite{fujii} & 3129 & 50655 \\
 $^{28}$Si-24Pr7 & 99.9994751(20) & 0.0004815(16) & 0.0000434(09) & Table 5 in Ref.\ \cite{fujii} & 9730 & 71244 \\
\hline
\end{tabular}
\end{center}
\caption{\small Comparison of the isotopic compositions for several \esi-enriched crystals (bottom four rows) to the natural abundance of stable silicon isotopes (uppermost row), listed with uncertainties quoted in parentheses and corresponding to the last two significant figures. In the two rightmost columns, we calculate the post-enrichment reduction factors for $^{29}$Si and $^{30}$Si relative to the corresponding IUPAC natural abundance.}
\label{tab:silicon-28}
\end{table*}

Over the last 25 years a research program --- often collectively referred to as the ``Avogadro Project'' --- has focused on redefining the kilogram using monocrystalline silicon that is highly enriched in  \esi\ \cite{seyfried,jertz, becker,friedrich,becker2,fujii},  encompassing a variety of materials and metrological techniques for achieving an improved, lower-uncertainty definition of the kilogram mass. The use of a single crystal of \esi with ultra-high purity, both chemically and isotopically, has the following benefits for metrology:
\begin{enumerate}[i)]
\item Minimizing chemical impurities reduces uncertainty in lattice defects and total crystal mass \cite{Agostino};
\item Minimal lattice defects and the physical regularity of the crystal structure allow for high-precision atom counting \cite{fujii}; and
\item Having high isotopic purity (and thus a single atomic mass) further reduces uncertainty when estimating the total mass of an enriched crystal \cite{pramann}.
\end{enumerate}

This research program is made possible via gaseous ultracentrifugation followed by chemical reduction to metal \cite{becker3,fujii}. The enrichment process begins with Na$_2$$^{\textrm{nat}}$SiF$_6$ conversion to $^{\textrm{nat}}$SiF$_4$ gas. After gaseous ultracentrifugation enrichment, the $^{28}$SiF$_4$ is collected and prepared for further refinement by conversion to silane gas ($^{28}$SiH$_4$) which is subjected to cryofiltration and rectification to remove chemical impurities. The chemically purified silane gas is then converted to metal through thermal decomposition onto a slim \esi rod, producing polysilicon enriched in \esi. 
Finally, a single crystal is grown using multiple FZ runs (to further increase purity) and is manufactured into $\sim$1\,kg spheres.

The process described above has been systematically improved over several production campaigns, resulting in extraordinary chemical and isotopic purities. Table \ref{tab:silicon-28} reproduces the reported isotopic-purity levels obtained in the various development stages of this program. Of particular interest are the reduction factors (from stable-isotope abundance levels) in the isotopic concentration fractions of $^{29}$Si and $^{30}$Si in the post-enrichment crystals, presented in the rightmost columns. It is plausible that the corresponding \si reduction factors are at least as large as for $^{30}$Si (if not larger). Making this assumption, a simple calculation suggests that the \si activity of 80 decays/(kg\,d) measured by the DAMIC CCDs could be reduced to a residual level of $\sim$0.001 decays/(kg\,d) if the CCDs were to be fabricated from isotopically pure \esi, such as the material identified as \esi-24Pr7 in Table \ref{tab:silicon-28}.

We have not found in the literature any reporting of \si levels for these enriched crystals. Taking again the \si activity measured by the DAMIC CCDs, and considering the 153-year half-life, 1 kg of detector material would contain only 6.4$\times$10$^{6}$ \si atoms. In terms of an isotopic concentration fraction, this is equivalent to 3$\times$10$^{-19}$. The $^{30}$Si fractions reported in Ref.\ \cite{fujii} (and listed in Table \ref{tab:silicon-28}) have uncertainties that suggest a limitation in measurement sensitivity at the level of 9$\times$10$^{-9}$. This level of sensitivity was achieved using a highly sophisticated mass-spectrometry technique~\cite{pramann2},
and it falls short of the \si concentration in the DAMIC CCDs by ten orders of magnitude. It is 25 orders of magnitude away from the \si level inferred above for \esi-enriched material. We conclude that new methods for evaluating \si  levels throughout the silicon production cycle --- commercial or enriched --- are required to qualify material for future silicon-based dark matter detectors.

\section{Assay Concepts}\label{sec:assay}
At present, the measurement of \si by the DAMIC CCDs~\cite{damic_backgrounds} is the only proven method with sufficient sensitivity to perform meaningful \si assay in silicon metal. Fabrication of single-crystal silicon wafers into DAMIC-style CCDs, followed by operation in a sufficiently low-background installation, is a viable assay technique for achieving relevant and useful low-level sensitivity to \si decays in end-product devices.  Additional measurements of this type are of value to rigorously confirm the level of \si seen in CCDs, to either reduce the uncertainty in the activity level per kg of silicon metal or to explore the batch to batch \si variability (or both). 

As noted in the previous section, to evaluate \si levels throughout the silicon production cycle (i.e., up-stream of device fabrication) requires development of new assay methods. The extremely low levels of \si in silicon metal make radiometric counting of the daughter \p decay an attractive possibility. In the specific case of the DAMIC CCDs in Ref.~\cite{damic_backgrounds}, there were only $\sim$10$^6$--10$^7$ \si atoms available for measurement per kg of silicon material. If separated from the bulk silicon matrix, the \p daughter provides a viable radiometric measurement using a low-background beta-detection gas cell \cite{bids1,bids2}.  Thus, it is the extraction of the \p atoms from the bulk silicon sample that is crucial for establishing a low-level \si assay method.

Low-level \si assay has been developed as an age-dating tool for measuring and understanding sedimentation in support of geochronology research. This geochronological tool is designed to bridge the gap between $^{210}$Pb (half-life $<$\,100 years) and $^{14}$C (half-life $>$\,1000 years). As discussed in Sec.\ \ref{sec:environment} and suggested in Fig.\ \ref{fig:silicon-32}, \si is deposited into the terrestrial environment from the atmosphere and then incorporated into algae diatoms as biogenic silica where it becomes a constituent of sediments after the organisms die. Chemical-separation schemes have been developed for processing large masses of bulk marine and freshwater sediments for isolating biogenic silica and extracting the  \p daughter in a purified form that is then used for radiometric counting \cite{fifield,morgenstern2}. These methods have facilitated reconstruction of sedimentation records in order to study environmental conditions and human impacts on different earth systems.

We describe here the sample-preparation process used for geochronology measurements that was developed at Pacific Northwest National Laboratory (PNNL) for processing marine sediments from the Puget Sound to infer extant \si levels. The available literature was utilized for initial development and modifications were made where necessary \cite{fifield,morgenstern2,demaster,benitez,conley}. After an initial refinement via a physical-separation process, steps are taken to selectively leach the biogenic silica in weak sodium hydroxide and then precipitate the purified silica in nitric acid. The target \p atoms, previously in secular equilibrium with the \si, and any naturally occurring (non-silicon) radiogenic impurities stay in solution, leaving a cleanly separated silica gel that is rinsed and stored. After a couple months the \p will have grown back into secular equilibrium in the separated silica gel. The gel is then re-dissolved in sodium hydroxide and precipitated once again to collect the phosphorous. This step is known as phosphorus ``milking'' and requires retention of the supernatant that contains the \p analyte of interest. The silica gel can be retained for additional milking cycles and subsequent repeat measurements. After milking, the phosphorus solution is chemically pure but has not yet reached the radiopurity required for low-level beta counting; two additional precipitation steps are carried out, as well as a final cation exchange procedure. Finally, the liquid sample is reduced in volume and applied to a piece of filter paper that is then sandwiched between thin films and mounted inside a beta-detection gas cell (as illustrated in Fig.~5 of Ref.~\cite{bids1}).

Although originally developed for geochronology, this assay method is potentially useful for analysis of materials in the silicon production cycle to help ensure growth of crystals that have low levels of \si. However, the method requires adaptation for use with different forms of silicon such as solid metal (vs.\ marine sediments). In light of the discussion of isotopically pure \esi in Sec.\ \ref{sec:isotopically_pure}, it is also worth considering starting with samples of silicon in a gaseous form. In both cases the objective is to develop methods for \p extraction so as to enable radiometric counting. We propose and review \p extraction methods in the following Secs.\ \ref{ssec:solid}--\ref{ssec:p_removal}, followed by a discussion in Sec.\ \ref{ssec:sampling} of potential assay sampling points throughout the traditional production cycle and the alternative \esi-enrichment process. We conclude in Sec.\ \ref{ssec:survey} by emphasizing the importance of using assay to more thoroughly evaluate \si levels in silicon from existing sources and producers.

\subsection{Dissolution of solid silicon metal}\label{ssec:solid}
The PNNL process for \si assay of marine sediments was designed to purify silicates (SiO$_2$), which readily dissolve in basic solutions. Dissolution of monocrystalline silicon is more challenging as it is extremely inert, and thus front-end modifications of the method are required.  The least troublesome dissolution utilizes a mixture of nitric acid and hydrofluoric acid in a 1:1 ratio. This mixture is commonly used in the silicon-wafer industry  to etch away contaminants \cite{kant}. For large masses, as is needed to accumulate a sufficient number of daughter \p atoms, it is advantageous to mill or brake the silicon into small pieces prior to dissolution in order to increase the total surface area, thereby reducing the time required for complete dissolution.

The dissolution chemistry itself is rather complex due to the reactivity of silicates with fluoride species. Two main chemical reactions take place, with the first being the reaction of silicon with HNO$_3$ to produce SiO$_2$, NO, and H$_2$O. The second is the decomposition of SiO$_2$ by HF to form hexafluorosilicic acid and more water \cite{steinert, schwartz}. The representative individual chemical reactions are
\begin{align}
3\textrm{Si} + 4\textrm{HNO}_{3} & \rightarrow 3\textrm{SiO}_{2} + 4\textrm{NO} + 2\textrm{H}_{2}\textrm{O} \\
\textrm{SiO}_{2} + 6\textrm{HF} & \rightarrow \textrm{H}_{2}\textrm{SiF}_{6} + 2\textrm{H}_{2}\textrm{O}
\label{eq:1}
\end{align}
with a combined total dissolution reaction of
\begin{align}
3\textrm{Si} + 4\textrm{HNO}_{3}+ 18\textrm{HF} & \rightarrow 3\textrm{H}_{2}\textrm{SiF}_{6} + 4\textrm{NO} + 8\textrm{H}_{2}\textrm{O}.
\label{eq:2}
\end{align}
As an exothermic reaction, other products are created and gases are evolved without application of additional heat, including those that would result in losses of silicon from the system (e.g.\ silicon tetrafluoride and volatilized hexafluorosilicic acid); note that the secondary reaction causes production of silicon tetrafluoride gas via the reaction
\begin{align}
\textrm{Si} + 4\textrm{HF} & \rightarrow  \textrm{SiF}_{4} + 2\textrm{H}_{2}.
\label{eq:3}
\end{align}
After total dissolution is achieved, the silicon can be fully volatilized by simultaneously heating while adding more HF, leaving behind (in principle) any \p in the form of residual solid compounds and thus dramatically reducing the effective mass of the sample.  At this point, the phosphorous can be put back into solution and then purified and detected as in the PNNL method described in the previous section. The full process requires evaluation of chemical yields and characterization of any radioactive backgrounds (introduced by the process) that might affect the measurement sensitivity for low-background counting of \p betas.

\begin{figure*}[ht]
\centering
\includegraphics[width=0.95\textwidth]{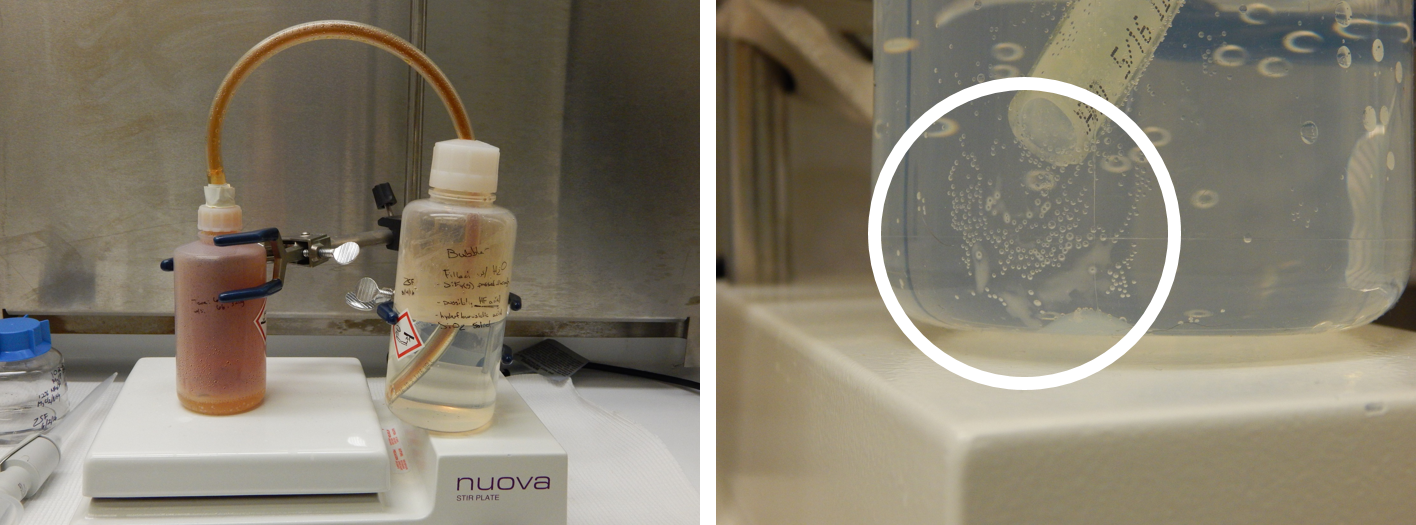}
\caption{\small Feasibility test of chemical dissolution of solid silicon metal for introduction into the \si radiometric-assay methodology outlined in the main text. Left: Reaction vessel (darker color) with gases bubbled through water vessel. Right: Formation of SiO$_2$ gel on side of bubbler vessel.} 
\label{fig:dissolution}
\end{figure*}

In the PNNL method, there are additional chemical-purification steps because the matrix for environmental samples is complicated. Pure silicon metal may not have this issue, depending on the sampling location in the silicon supply chain. This suggests the possibility of assay for \p via a one-step dissolution process, in which the sample is immediately reduced in volume and loaded for radiometric counting after the silicon is fully volatilized. As a benefit, such a method foregos the use of additional chemicals that could introduce trace levels of radioactive contamination. However, volatilization of silicon without reclamation prohibits multiple measurements (via phosphorous re-milking).  It would also result in loss of potentially expensive material if applied to the assay of samples enriched in \esi. In either case, large silicon masses (up to 1 kg) must be processed in order to extract a sufficient number of \p atoms, which will require multiple liters of high-purity acid and large digestion systems (e.g.\ HF-resistant fume hoods and labware).

While the use of large volumes of acid is unavoidable due to the stoichiometry of the reactions, a potential solution for silicon reclamation is to utilize the hydrolysis of gaseous  species to produce SiO$_2$ as a gel:
\begin{align}
\textrm{SiF}_{4} + 2\textrm{H}_{2}\textrm{O} & \rightarrow  \textrm{SiO}_{2} + 2\textrm{H}_{2}\textrm{SiF}_{6}.
\label{eq:4}
\end{align}
For example, the dissolution system could contain a reaction vessel with gases forced through a bubbler system where they would react with a stable aqueous medium (i.e.\ purified water), such that the silicon product would be left behind for future processing. This apparatus would need to be designed and optimized around a suite of parameters to achieve high efficiency while maintaining purity.  Of particular importance is the need for assiduously clean wet-chemistry protocols to prevent introduction of beta-emitting environmental backgrounds (e.g.\ $^{210}$Pb, $^{40}$K and $^{14}$C). Further, care must be taken to ensure that all reagents and water are initially devoid of dissolved silicon (and thus \si). Figure \ref{fig:dissolution} illustrates the execution of a small-scale experiment to demonstrate the feasibility of the concept.

\subsection{Separation of P-32 via gaseous distillation}\label{ssec:distillation}
\begin{table}[!b]
\small
\begin{center}
\begin{tabular}{ c c r r r}
\hline 
& & & & \\[-1em]
 &  & \multicolumn{2}{c}{\underline{Boiling Point, T$_{\textrm{bp}}$ ($^{\circ}$C)}} & $\Delta$T$_{\textrm{bp}}$\\
& & & & \\[-1em]
Source & Analogue & Si species & P species &  ($^{\circ}$C)\,\,\\
& & & & \\[-1em]
\hline
& & & & \\[-1em]
SiF$_4$ & PF$_{3}$ & $-86.2$~~ & $-101.6$~ & $-15.4$\,\\
SiF$_4$ & PF$_{5}$ & $-86.2$~~ & $-84.9$~ & $1.3$\,\\
SiH$_4$ & PH$_{3}$ & $-111.5$~~ & $-87.5$~ & $24.0$\,\\
HSiCl$_3$ & PCl$_{3}$ & $31.9$~~ & $74.2$~ & $44.3$\,\\
& & & & \\[-1em]
\hline
\end{tabular}
\end{center}
\caption{\small List of silicon source gases and potential \p-containing analogue species that may exist as a product of the \si beta decay. Boiling points (from Ref.\ \cite{boiling_point}) of both the Si and P species are listed as well as the boiling-point difference $\Delta$T$_{\textrm{bp}}$. SiH$_4$ and HSiCl$_3$ correspond to stages within the commercial silicon production cycle following chemical purification but prior to formation of solid polysilicon metal.  Similarly, SiF$_4$ and SiH$_4$ correspond to isotopically and chemically purified gases in the production cycle used by the Avogadro Project (see Sec.~\ref{sec:isotopically_pure}) prior to formation of \esi-enriched polysilicon metal.}
\label{tab:boiling_points}
\end{table}

Assay of precursor gases (e.g.\ SiF$_4$, SiH$_4$, HSiCl$_3$, etc.)\ may be more straightforward and less prone to introduction of sensitivity-limiting radiocontaminants. Hypothetically, it may be possible to separate the \p from a precursor gas via distillation. This assumes that the \p, following the \si beta decay, is in a form that can be separated and/or distilled from the source gas. Table \ref{tab:boiling_points} lists typical precursor gases and potential \p analogue species that may exist after molecular rearrangements and/or reactions, as well as their respective boiling points. In principle, a separation method could be developed to take advantage of the difference in boiling point, perhaps via cryogenic capture of the \p species on a target (e.g.\ filter media) or by chemical conversion in a reaction vessel (``gettering''). The isolated and concentrated sample could then be loaded (as a gas) into a low-background gas proportional counter for detection of the \p betas (using, e.g., the counters described in Ref.\ \cite{ulbpc}). This approach has the advantage of involving limited handling and reagents, which should reduce the potential for introduction of the type of environmental radiocontaminants that will be a persistent challenge in the wet-chemistry methods described in Sec.~\ref{ssec:solid}. However, significant development work would be required to investigate  optimum distillation protocols, as well as procedures for effectively determining capture and detection efficiencies.

\subsection{Additional phosphorous collection methods}\label{ssec:p_removal}
In the following subsections, we review additional established methods from the literature for isolating phosphorous from silicon materials.  In both cases, the techniques were not developed specifically with \p in mind. Consequently, similar to the concepts presented in Secs.\,\ref{ssec:solid} and \ref{ssec:distillation}, significant development work would be required to characterize \p separation at a scale that could be measured radiometrically.

\subsubsection{Removal from trichlorosilane}\label{sssec:trichlorosilane}
As noted in Sec.\ \ref{ssec:production-cycle}, trichlorosilane is the primary precursor in the production of most high-purity polysilicon.  A laboratory assay method to determine the total amount of PCl$_3$ in liquid trichlorosilane suggests a route for concentration of \p from the bulk material.  The method complexes PCl$_3$ with CuCl and ethanol.  The trichlorosilane is then removed via evaporation.  The non-volatile residue is dissolved in hot nitric acid and prepared for mass spectrometry.  A recovery efficiency of 100\% for a PCl$_3$ spike has been demonstrated using electrothermal vaporization ICP-MS with a detection limit of 0.02 ng/g \cite{wei}. This level of sensitivity is still many orders of magnitude away from the \si level measured by the DAMIC CCDs.  However, the published procedure mixed the sample and reagents for only 3 hours; if the kinetics are reasonable, it may be possible to convert this to a quasi-continuous process in which a larger amount of trichlorosilane is slowly mixed with the CuCl and ethanol near the boiling point over a period of several days.  The evaporated trichlorosilane could be recovered and stored to allow the \p to re-accumulate, thus making it possible to perform multiple measurements. Detection of \p in the residue via mass spectrometry would likely still fall short of the requisite sensitivity due to the presence of stable phosphorous isotopes and/or other similar-mass interferences.  Consequently, additional method development would be needed to prepare the residue for radiometric analysis.

\subsubsection{Removal from metallic silicon}\label{sssec:metalic_si}
Increased demand for solar-grade silicon has motivated research into using metallurgical processing methods to upgrade metallurgical-grade silicon for production of solar cells \cite{Safarian}, resulting in new methods for removal of volatile impurities. The vapor pressure of phosphorous is relatively high and its segregation coefficient in silicon is low ($k=0.35$); so phosphorous impurities will tend toward the liquid phase.  Directional solidification of silicon therefore drives phosphorous into the melt with subsequent evaporation at the surface under vacuum conditions.  Phosphorous diffusion through the melt is a first order effect.  Consequently, directional solidification of molten silicon under high vacuum reduces the total phosphorous concentration of the bulk material with a significant contribution from surface evaporation \cite{Jiang}. The combination of vacuum refining and adding sodium- or calcium-rich slag layers was shown to be effective in reducing phosphorous from $\sim$20 to 1 ppm \cite{Huang}.  Although effective, it is unclear if such methods could be used to recover \p for radiometric analysis.

\subsection{Potential assay sampling points}\label{ssec:sampling}
There are several locations within the commercial silicon production cycle where testing for the presence of \si may be appropriate:
\begin{enumerate}[i)]
\item Raw geologically sourced silicon ore,
\item Silicon materials at the entrance to or immediately after polysilicon production,
\item Siemens or fluidized bed reactors where chlorine is present,
\item Aqueous solutions used during production of high-purity silicon crystals, or
\item Final-product semiconductor-grade or single-crystal silicon metal.
\end{enumerate}
These sample locations, and the form of the silicon material at each stage, mirror the \si injection vectors outlined in Secs.\ \ref{ssec:vector1}--\ref{ssec:vector5}. In principle, with sufficiently sensitive analytic methods for performing \si assays, it would be possible to identify the most important vectors and thereby perform a \si-controlled production of silicon substrates for use as low-background radiation detectors.

It is also worth considering the case of the isotopically pure \esi discussed in Sec.\ \ref{sec:isotopically_pure}, which suggests analogous sampling locations.  Of particular interest would be gaseous samples prior and subsequent to the enrichment process to validate the effectiveness of enrichment for reduction of \si. Naively, we expect that the level of \si in enriched material is so low that it will be undetectable, even with any increased measurement sensitivity obtained through the assay-method development suggested in Secs.\ \ref{ssec:solid}--\ref{ssec:p_removal}. Consequently, enriched SiF$_4$ gas or single-crystal silicon would likely serve as a valuable process ``blank'' for understanding assay-method detection limits.  However, it is still possible that \si is introduced into enriched material subsequent to the enrichment process, highlighting the importance of aqueous processing solutions as \si assay targets.

\subsection{Silicon source material evaluations}\label{ssec:survey}
  The prior sections' descriptions of assay methods address some of the tools that may be needed to conduct a more complete evaluation of \si in source materials used for the commercial production of silicon metal. Further, we believe that a comprehensive program of \si measurements for a variety of silicon sources is a requisite step in formulating an effective \si-mitigation strategy (outlined in the following section). For example, it would be instructive to work with single-crystal vendors to select a representative set of wafer samples that could be fabricated into DAMIC-style CCDs in order to help identify any existing ``off-the-shelf'' sources of single-crystal silicon with acceptably low levels of \si. More broadly, a world survey of silicon sources might entail measuring samples from quarry mines, high-purity quartz deposits, and beaches, some of which could be selected to span a range of environmental factors (e.g., low- versus high-precipitation geographies).\footnote{It is important to note that silicon is sold globally as a commodity gas. Consequently, the high-purity polysilicon feedstock used by single-crystal growers may contain silicon from multiple geologic sources.}

\section{Implications for dark matter research}\label{sec:implications}
In this section, we evaluate potential mitigation measures targeted at creating future silicon detectors with low levels of \si, under the assumption that a silicon source with a sufficiently low level of \si is not revealed by the survey program suggested in Sec.~\ref{ssec:survey}. 

To provide a target for an allowable residual level of \si in substrates used as dark matter detectors, we first compare the rate of \si (and daughter \p) decays in such a detector to the coherent-scattering rate of solar neutrinos with nuclei. Coherent scattering of solar neutrinos with nuclei is a fundamentally limiting background process to the measurement of dark matter interactions. Thus, trace \si contamination is no longer a concern if the residual \si activity corresponds to a rate below the solar-neutrino coherent-scattering rate.

\begin{figure}[!t]
\centering
\includegraphics[width=\textwidth]{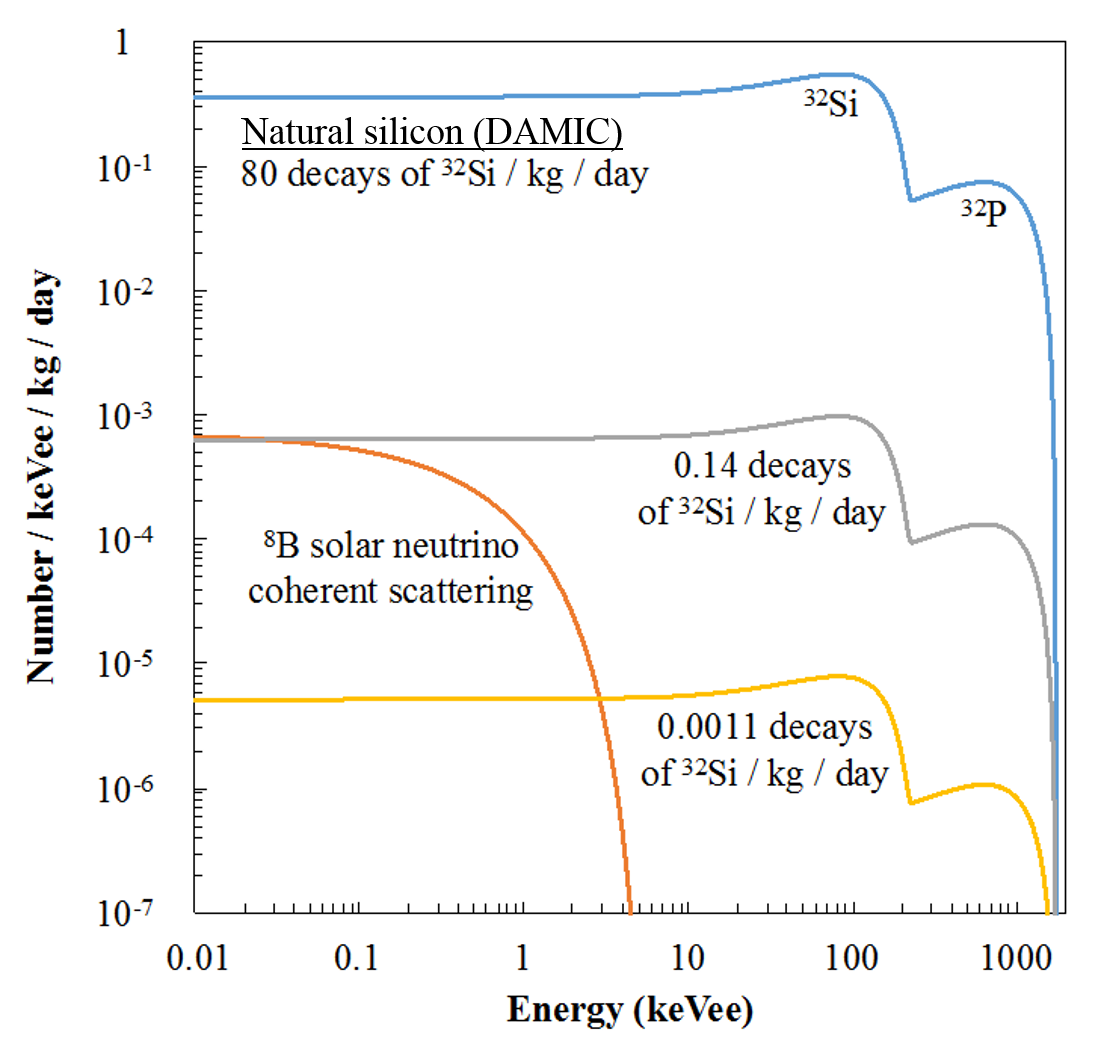}
\caption{\small Si-32 plus \p beta-decay spectra normalized to various \si levels (noted in terms of \si decays/kg/day) in a silicon detector compared to the rate of nuclear recoils induced by coherent scattering of $^8$B solar neutrinos (orange featureless curve). The \si and daughter \p beta-decay spectra are shown summed together (227.2 and 1710.7 keV endpoints, respectively). The uppermost curve is representative of the \si level reported by DAMIC \cite{damic_backgrounds}, whereas the middle \si-\p decay spectrum is normalized such that the rate matches the $^8$B solar-neutrino scattering rate at zero recoil energy. The lowest \si-\p decay spectrum assumes reduction of \si (relative to the DAMIC level) equal to the $^{30}$Si isotopic reduction factor (1/71244) indicated in Table \ref{tab:silicon-28} for the enriched crystal identified as Si28-24Pr7.} 
\label{fig:floor}
\end{figure}

For simplicity, we assume that our hypothetical detector is composed of 100\% \esi\footnote{This assumption of pure \esi relates to the calculation of the coherent-scattering rate for $^{8}$B solar neutrinos and as such has no bearing on the \si activities assumed in Fig.\ \ref{fig:floor}. }, and we focus solely on $^8$B solar neutrinos. Figure \ref{fig:floor} shows the coherent-scattering spectrum from $^8$B neutrinos compared with the combined spectrum of \si and \p beta decays, with the latter normalized to three different levels of \si activity and with \p in secular equilibrium. In our calculation of the $^8$B spectrum, we have used the total $^8$B solar-neutrino flux of 5.25$\times$10$^6$ cm$^{-2}$s$^{-1}$ \cite{sno},  a readily tabulated $^8$B solar-neutrino spectral shape \cite{shape}, the neutrino coherent-scattering cross section \cite{cs1,cs2,cs3}, a simple formulation of the weak nuclear charge \cite{weak_charge}, and the Helm form factor \cite{helm}. Other input data include values for several physical constants --- the Fermi coupling constant ($G_{F}$), the weak-mixing angle (sin$^2$$\theta_{W}$), the unified atomic mass unit ($u$), and the Planck constant ($\hbar$$c$) \cite{pdg} --- and data on the properties of silicon nuclei \cite{snuc1,snuc2}. Beta-decay spectral shapes for \si and \p are from the \textsc{BetaShape} program \cite{betashape,betashape2} 
using isotope data and $\log ft$ values from Ref.\ \cite{nndc_page}. To convert the $^8$B solar-neutrino nuclear-recoil energy spectrum to an equivalent electron-recoil energy scale (i.e.\ units of ``keVee''), as is appropriate for the \si and \p beta-decay spectra, we have used the Lindhard model \cite{lindhard} as characterized in Ref.\ \cite{mei_lindhard}.\footnote{Although it does not significantly alter our conclusions, it is important to note that the Lindhard model is an approximation that has been shown to deviate from experimental results for energies below 0.4\,keVee in silicon~\cite{damic_photoneutron_PRD,photoneutron_jinst}.}

Figure \ref{fig:floor} clearly demonstrates the potential benefits of \si mitigation for dark matter research. The information presented in this report suggests two principal routes to producing silicon detectors with such low levels of \si: (1)\,Sourcing naturally occurring silicon from geologic formations that are likely to be low in \si and shepherding the material through refinement and crystal production so as to prevent \si introduction; and (2)\,Use of \esi-enriched material and a single-crystal production process such as demonstrated by the Avogadro Project. Both are discussed and evaluated below.

\subsection{Mitigation method 1: sourcing low-Si-32 ore}\label{ssec:mitigation1}
As illustrated in Fig.\ \ref{fig:silicon-32}, \si is initially introduced from the atmosphere via precipitation; so it is reasonable to assume that there is a general and uniform level present worldwide if averaged across the Earth's surface on a large enough scale. Localized \si levels, however, may depend on a number of smaller-scale effects and might therefore vary significantly.  For example, silicon deposits with minimal exposure to surface waters may have lower \si than deposits mined from or near surface-water runoff paths. Additionally, the local microclimate may affect the transport of \si from the atmosphere and into the strata of quartzite deposits simply due to seasonal variations or differences in total annual precipitation levels \cite{MORGENSTERN3}. Consequently, silicon ore that has been isolated from surface waters for $\sim$1500 years (or more) is likely needed to ensure significantly lower \si levels and thus to make it a suitable raw material for the production of future dark matter detectors. This suggests the need to develop high-sensitivity assay methods in order to verify that the source material is indeed sufficiently low in \si (see Sec.\ \ref{sec:assay}).

Further, the low-\si ore must be refined separately from other silicon sources, and any water used in the refining process must also have low concentrations of \si. Again, the assay methods suggested in Sec.\ \ref{sec:assay} could evaluate the refinement and production processes to ensure little additional \si is introduced into the material targeted for fabrication of dark matter detectors. As noted in the \esi-enrichment literature \cite{inkret}, it is also important to control use of silicon-containing laboratory equipment because of the potential for isotopic contamination. 

Although finding a natural source of silicon that has a low level of \si may be feasible, neither has such a source been identified nor is the crystal-production process that can maintain the isotopic purity readily available. The assay methodology discussed in Sec.\ \ref{sec:assay} may provide a means to address these two initial issues in establishing this \si mitigation method.

\subsection{Mitigation method 2: using isotopically pure silicon}\label{ssec:mitigation2}
The Avogadro Project has demonstrated a complete production chain for single-crystal silicon that has a high chemical purity and is isotopically enriched in \esi. We presume that this enriched material is adequate, or could be made adequate with little additional processing, for use as substrate material for silicon-based dark matter detectors. If the $^{30}$Si isotopic reduction factor is representative of a corresponding reduction in \si, Fig.\ \ref{fig:floor} shows that the corresponding rate of \si and \p beta decays would be sub-dominant to the coherent-scattering rate of solar neutrinos, and thus the level of \si in this enriched material can be sufficiently low for use in future dark matter detectors.

There are two key uncertainties associated with this mitigation strategy: cost and the \esi production rate. Future silicon-based experiments will require tens of kilograms of active detector material to directly detect the dark-matter--nucleon cross section at a level corresponding to the rate at which solar neutrinos become the limiting background. The literature associated with the \esi-enrichment program indicates recent production rates that are generally on the scale of tens of grams, with a couple examples of production of $\sim$5 kg quantities (cf.\ Table 7 in Ref.\ \cite{wang}).

It is noteworthy that the literature includes examples of production of tens of grams of $^{29}$Si at isotopic-purity levels $>$\,99.9\% \cite{wang,Churbanov}.  This is interesting from the perspective of investigating the spin dependence of the interaction of dark matter with detector materials. The \esi nucleus has no nuclear spin (0$^{+}$), whereas $^{29}$Si has a non-zero nuclear spin ($\frac{1}{2}^{+}$). This suggests the possibility of producing detectors from both \esi- and $^{29}$Si-enriched crystals to test the spin-independent and -dependent couplings of dark matter to normal matter, respectively. However, as the natural abundance of $^{29}$Si is a factor of 20$\times$ lower than \esi, one might reasonably anticipate at least an order of magnitude increase in cost associated with obtaining a comparable mass of isotopically pure $^{29}$Si relative to similar-purity \esi.

\section{Conclusions}\label{sec:conclusions}
The first definitive measurement of \si in detector-grade silicon \cite{damic_backgrounds} has propelled it forward as a particular background concern for future  dark matter direct-detection experiments. We have reviewed the origin of \si and its subsequent transport through the environment and the commercial silicon production cycle, in particular as pertains to the introduction of \si into material used to fabricate radiation detectors. We also identified silicon enriched in \esi as a potential alternative to the commercial supply chain. We have outlined methods for assaying silicon materials --- before, during, and after silicon refinement and single-crystal growth --- to evaluate and ensure selection of silicon substrates that have sufficiently low levels of \si for use in future silicon-based dark matter detectors. Additionally, we have suggested methods for mitigation of \si, guided by a target residual \si level that would result in a background rate that is sub-dominant to the coherent-scattering rate of $^{8}$B solar neutrinos. Use of isotopically pure \esi appears favorable as a mitigation strategy if the issues of cost and scale can be addressed.  

\section*{Acknowledgement}
Pacific Northwest National Laboratory (PNNL) is operated by Battelle for the United States Department of Energy (DOE) under Contract No.\ DE-AC05-76RL01830. The laboratory investigation into direct methods for \si assay in silicon metal was supported by the DOE Office of High Energy Physics' Advanced Technology R\&D subprogram, and the evaluation of the transport of \si into silicon detectors and its potential impact on future dark matter direct-detection experiments was supported under the Nuclear-physics, Particle-physics, Astrophysics, and Cosmology (NPAC) Initiative, a Laboratory Directed Research and Development (LDRD) effort at PNNL.  We thank Nathan Johnson at PNNL for development of Figures \ref{fig:silicon-32} and \ref{fig:production}, and we thank Jared Yamaoka for processing of the \si production rate for the chlorine cosmogenic-production channel using \textsc{Geant4}. Additionally, we thank Priscilla Cushman at the University of Minnesota for her initial investigations of the polysilicon refinement process and for suggesting the possibility of chlorine as a \si injection vector.

\section*{References}
\bibliography{references}

\end{document}